\documentclass[11pt,a4paper]{article}
\usepackage{jheppub_kim}

\usepackage{subfigure,amsmath,amssymb ,amsfonts, latexsym}

\usepackage[notcite,color,final
                              ]{showkeys}
\definecolor{refkey}{gray}{.25}
\definecolor{labelkey}{gray}{.25}

\newcommand{\ud}{\mathrm{d}}
\def \pd {\partial}

\newcommand{\be}{\begin{equation}}
\newcommand{\ee}{\end{equation}}
\newcommand{\beqa}{\begin{subequations}\begin{eqnarray}}
\newcommand{\eeqa}{\end{eqnarray}\end{subequations}}

\usepackage{color}

\begin{document}

\title{Bounce and cyclic cosmology in extended nonlinear massive gravity}

\author[a,b]{Yi-Fu Cai}  
\author[c]{Caixia Gao} 
\author[d,e,f]{Emmanuel N. Saridakis}

\affiliation[a]{Department of Physics, Arizona State University, Tempe, AZ
85287, USA}

\affiliation[b]{Department of Physics, McGill University, Montr\'eal, QC, H3A 2T8, Canada}

\affiliation[c]{Department of Physics, University of Mississippi, Oxford,
MS 38677, USA}

\affiliation[d]{Physics Division, National Technical University of Athens,
15780 Zografou Campus, Athens, Greece}

\affiliation[e]{CASPER, Physics Department, Baylor University,
Waco, TX  76798-7310, USA}

\affiliation[f]{Institut d'Astrophysique de Paris, UMR 7095-CNRS,
Universit\'e Pierre \& Marie Curie, 98bis boulevard Arago, 75014 Paris,
France}

\emailAdd{ycai21@asu.edu}
\emailAdd{cgao1@go.olemiss.edu}
\emailAdd{Emmanuel$_-$Saridakis@baylor.edu}




\abstract{
We investigate non-singular bounce and cyclic cosmological
evolutions in a universe governed by the extended nonlinear
massive gravity, in which the graviton mass is promoted to a scalar-field
potential. The extra freedom of the theory can lead to certain energy
conditions violations and drive cyclicity with two different mechanisms:
either with a suitably chosen scalar-field potential under a given
St\"{u}ckelberg-scalar function, or with a suitably chosen
St\"{u}ckelberg-scalar function under a given scalar-field potential.
Our analysis shows that extended
nonlinear massive gravity can alter significantly the evolution of the
universe at both early and late times.
}


\maketitle

\section{Introduction}

The question on whether there exits a consistent covariant theory for
massive gravity, where the graviton acquires a mass and leads to a
modification of General Relativity at large distances, was initiated by
Fierz and Pauli a long time ago \cite{Fierz:1939ix}. However, it was
observed that the nonlinear terms required by massive gravity \cite{Vainshtein:1972sx},
which can give rise to continuity of observables \cite{vdam, vdam2},
lead inevitably to the existence of the Boulware-Deser (BD)
ghost \cite{Boulware:1973my}, and thus make the theory unstable.

Although for the subsequent decades it was believed that there is no
consistent way to construct a massive gravity free of ghosts, a nonlinear extension of
massive gravity was constructed recently by de Rham, Gabadadze and
Tolley \cite{deRham:2010ik, deRham:2010kj}. In this model, the BD ghost can be
removed in the decoupling limit to all orders in perturbation theory
through a systematic construction of a covariant nonlinear action (see
\cite{Hinterbichler:2011tt} for a review). Although it is still
controversial
to prove the absence of BD ghost at the non-perturbative level, the theoretical
and phenomenological advantages led to a wide investigation of this theory
\cite{Koyama:2011yg, Hassan:2011hr, deRham:2011rn, CuadrosMelgar:2011yw,
D'Amico:2011jj, Hassan:2011zd, Kluson:2011qe, Gumrukcuoglu:2011ew,
Volkov:2011an, vonStrauss:2011mq, Comelli:2011zm, Hassan:2011ea,
Berezhiani:2011mt, Gumrukcuoglu:2011zh, Khosravi:2011zi, Brihaye:2011aa,
Park:2010rp,Park:2010zw,
Buchbinder:2012wb, Ahmedov:2012di, Bergshoeff:2012ud, Crisostomi:2012db,
Paulos:2012xe, Hassan:2012qv, Comelli:2012vz, Sbisa:2012zk, Kluson:2012wf,
Tasinato:2012mf, Morand:2012vx, Cardone:2012qq, Baccetti:2012bk,
Gratia:2012wt, Volkov:2012cf, DeFelice:2012mx, Gumrukcuoglu:2012aa,
deRham:2012kf, Berg:2012kn, D'Amico:2012pi, Baccetti:2012re,
Fasiello:2012rw, D'Amico:2012zv, Baccetti:2012ge,
Langlois:2012hk,Gong:2012yv}.

Despite the successes of nonlinear massive gravity, it was also
noticed that certain cosmological instabilities still exist in the case
where the physical and the fiducial metrics have simple homogeneous and
isotropic forms \cite{DeFelice:2012mx}. This behavior motivated
researches towards extensions of nonlinear gravity models, namely the
construction of nonlinear massive gravity with less symmetric metrics
\cite{D'Amico:2011jj, Gumrukcuoglu:2012aa}. However, in \cite{Huang:2012pe}
a different approach was followed, and nonlinear massive gravity was
extended allowing for the graviton mass to vary. This could be realized
by introducing an additional scalar field, which coupling to the graviton
potentials produces an effective, varying, graviton mass. Moreover, this
extension provides a natural way to modify General Relativity not only in
the IR but also in the UV, since the graviton mass can be evolved into a
large value at the early universe \cite{Saridakis:2012jy}. Therefore, it
is interesting to study the cosmological implications of this scenario at
early
times, and this is a main topic of the present work.

On the other hand, it is well known that cosmological evolution governed
by standard Einstein gravity usually suffers from the problem of initial
singularity if Null Energy Condition (NEC) is preserved \cite{Borde:1993xh}.
A potential solution to the cosmological singularity problem may be provided
by non-singular bouncing cosmologies
\cite{Mukhanov:1991zn, Brandenberger:1993ef, Cai:2012va}. Such scenarios
have been constructed within various approaches to modified gravity
\cite{Veneziano:1991ek, Khoury:2001wf, Brustein:1997cv,
Kehagias:1999vr, Shtanov:2002mb, Saridakis:2007cf,
Cai:2010zma, Cai:2011tc, HLbounce, Cai:2009in,Saridakis:2009bv,Leon:2009rc,
Bojowald:2001xe,
Martin:2003sf,Saridakis:2010mf,Biswas:2010zk,Leon:2010pu,Biswas:2011ar,
Biswas:2012bp}.
Generally, a non-singular
bouncing model
can be acquired by using NEC violating matter \cite{
Dabrowski:2003jm,Dabrowski:2004hx,Yifu1, Yifu1b, Yifu2,
Yifu2b, Cai:2008qw}, by making use of various mechanisms \cite{Khoury,
Creminelli, Chunshan, Tirtho1, Tirtho3, Taotao, Damien, Abramo:2007mp}.
Note that in the case of a positive curvature a generic bounce
can be obtained by violating Strong Energy Condition (SEC) only
\cite{Starobinskii:1978yy}, however the singularity reappears in the fact
that the number of regular bounces is finite \cite{Page:1984qt}.
Furthermore, the perturbation theory of non-singular bounce cosmology and
its relation to observables was extensively studied in the literature
\cite{Yifu2, Yifu2b, Cai:2008qw, Cai:2009hc, Cai:2009rd, Cai:2009fn,
Liu:2010fm, Cai:2011zx, Cai:2011ci}. The extension of all the above bouncing
scenarios is the old idea of cyclic cosmology \cite{tolman}, in which the
universe presents a periodic sequence of contractions and expansions.
Cyclic cosmology has attracted a significant interest the last years
\cite{Steinhardt:2001st} since it brings different insights for the origin
of the observable universe \cite{Lidsey:2004ef, Piao:2004hr, Piao:2004me, Xiong:2007cn,
Piao:2009ku, Piao:2010cf, Liu:2012gu, Xiong:2008ic, Cai:2009zp, cyclic,
cyclic1,Sahni:2012er}
(see \cite{Novello:2008ra, Cai:2011bs} for recent reviews).

In the present work we are interested in constructing scenarios of cyclic
cosmology in a universe governed by  the extended, varying-mass, nonlinear
massive gravity. Particularly, we first determine the
St\"{u}ckelberg-scalar function and we suitably reconstruct the form
of the potential of the scalar field  which leads to a cyclic universe.
Alternatively, we first determine the scalar potential and we reconstruct
the St\"{u}ckelberg-scalar function in order to obtain cyclicity.
Interestingly enough cyclicity is easily acquired in this framework, since
extended nonlinear massive gravity can violate certain energy conditions
and therefore has fruitful implications to physics of the universe at both
early and late times.

This paper is organized as follows. In Section \ref{model} we briefly
review the cosmological equations under the extended scenario of nonlinear
massive gravity with a scalar field being introduced to evolve the graviton
mass. In section \ref{bouncingsol} we construct scenarios of bouncing
and cyclic universe. In particular, in subsection \ref{caseb} we start
with a given St\"{u}ckelberg-scalar function and we reconstruct the
scalar potential; while in \ref{caseW} we start from a given scalar
potential and we determine the corresponding St\"{u}ckelberg-scalar
function that leads to cyclicity.
Finally, section \ref{Discussion} is devoted to the summary of our results.

\section{Cosmology in extended nonlinear massive gravity}
\label{model}

In this section we briefly review cosmology in extended nonlinear massive
gravity \cite{Huang:2012pe, Saridakis:2012jy}. In such a scenario the graviton
mass is upgraded into an evolving function depending on a cosmic scalar
field,
of which a canonical kinetic term and a standard potential are added in
the
action too. In this construction, the total action is written as
 \begin{eqnarray} \label{action0}
&& S= \int d^4x \sqrt{-g} \left[ \frac{M_P^2}{2} R + V(\psi) ( U_2 +
\alpha_3 U_3 + \alpha_4 U_4)
-  \frac{1}{2} \pd_\mu \psi \pd^\mu
\psi -W(\psi)
\right]+S_m,\ \ \
\end{eqnarray}
where $M_P$ is the reduced Planck mass, $R$ is the Ricci scalar, and $\psi$
is the newly introduced scalar field with $W(\psi)$   its usual potential
and $V(\psi)$ an additional potential coupling to the graviton potentials.
These
graviton potentials are given by,
\begin{align}
U_2  =  \mathcal{K}^\mu_{[\mu}\mathcal{K}^\nu_{\nu]} ~, \quad
U_3  = \mathcal{K}^\mu_{[\mu}\mathcal{K}^\nu_{\nu}\mathcal{K}^\rho_{\rho]}
~, \quad
U_4  =
\mathcal{K}^\mu_{[\mu}\mathcal{K}^\nu_{\nu}\mathcal{K}^\rho_{\rho}\mathcal{
K}^\sigma_{\sigma]} ~,
\end{align}
with
\begin{eqnarray}
\mathcal{K}^\mu_\nu \equiv \delta^\mu_\nu-\sqrt{g^{\mu\rho}f_{AB}\pd_\rho
\phi^A
\pd_\nu \phi^B } ~, \quad \mathcal{K}^\mu_{[\mu} \mathcal{K}^\nu_{\nu]}
\equiv \frac{1}{2}(\mathcal{K}^\mu_{\mu} \mathcal{K}^\nu_{\nu}
-\mathcal{K}^\mu_{\nu}\mathcal{K}^\nu_{\mu}) ~,
\end{eqnarray}
and similarly for the other antisymmetric expressions. Moreover,
$\alpha_3$ and $\alpha_4$ are dimensionless parameters. Additionally,
$f_{AB}$ is a fiducial metric, and the four-form fields $\phi^A(x)$
are the St\"{u}ckelberg scalars introduced to restore general
covariance \cite{ArkaniHamed:2002sp}. As it was shown in
\cite{Huang:2012pe}
 the above scenario is still free of the the BD ghost. Finally, in the
action (\ref{action0}) we can take into account the standard
matter action $S_m$, minimally-coupled to the dynamical metric,
corresponding to energy density $\rho_m$ and pressure $p_m$.

In order to derive explicitly the cosmological equations, we need to
impose certain ansatzes for the two metrics. For the physical metric we
consider a flat Friedmann-Robertson-Walker (FRW) form
(one can also straightforwardly investigate the non-flat case
\cite{Huang:2012pe}):
\begin{eqnarray}\label{ds2}
d^2 s = -N(\tau)^2 d \tau^2 +a(\tau)^2 \delta_{ij} d x^i d x^j ,
\end{eqnarray}
with $N(\tau)$ the lapse function and $a(\tau)$ the scale factor, and for
simplicity for the St\"{u}ckelberg fields we choose the forms as follows,
\begin{eqnarray}\label{phi0i}
  \phi^0 = b(\tau)  ,  ~~~~\phi^i =a_{ref} x^i,
\end{eqnarray}
with $a_{ref}$ a constant positive coefficient.
Note that although in standard massive gravity such a choice for the
dynamical metric cannot be accompanied by a simple choice for the fiducial
one, in the scenario at hand the extra freedom allows for a simple
Minkowski ansatz for the fiducial metric
 \begin{eqnarray}
f_{AB}=\eta_{AB}.
\end{eqnarray}

Variations of the action with respect to $N$ and $a$ give rise to
the following Friedmann equations
 \begin{eqnarray}
\label{Fr1}
3M_P^2 H^2& =& \rho_{MG}+\rho_m,\\
\label{Fr2}
-2 M_P^2 \dot{H}& =&\rho_{MG}+p_{MG}+\rho_m +p_m,
\end{eqnarray}
where we have defined the Hubble parameter $H=\dot{a}/a$, with
$\dot{a}=da/(Nd\tau)$. In the end we adopt $N=1$ for convenience. In the
above
expressions we have defined the energy density and pressure arising from
the modified gravitational sector as 
\begin{eqnarray}
\label{rhomg}
&&\rho_{MG} =\frac12 \dot{\psi}^2+W(\psi)+V(\psi)
\left(\frac{a_{ref} }{a}-1\right)[f_3(a) +f_1(a)] \ \ \  \ \\
\label{pmg}
&&p_{MG}  =\frac12 \dot{\psi}^2-W(\psi)-
V(\psi)f_4(a)-V(\psi)\dot{b}f_1(a) ~,
\end{eqnarray}
with
\begin{eqnarray}
&&f_1(a)
=3-\frac{2a_{ref}}{a}+\alpha_3\left(3-\frac{a_{ref}}{a}\right)\left(1-
\frac{a_{ref}}{a} \right)+\alpha_4\left(1-\frac{a_{ref}} {a}\right)^2
\nonumber \\
&&f_2(a) =
1-\frac{a_{ref}}{a}+\alpha_3\left(1-\frac{a_{ref}}{a}\right)^2+
\frac{\alpha_4}{3}
\left(1-\frac{a_{ref}}{a} \right)^3
 \nonumber\\
&&f_3(a) =3-\frac{a_{ref}}{a}+\alpha_3\left(1-\frac{a_{ref}}{a}\right)
\nonumber\\
&&f_4(a) = -\left[6-\frac{6a_{ref}}{a}+\left(\frac{a_{ref}}{a}\right)^2+
\alpha_3\left(1-\frac{a_{ref}}{a}\right)\left(4-\frac{2a_{ref}}{a}
\right) +\alpha_4\left(1-\frac{a_{ref}
}{a}\right)^2\right].\ \ \ \ \ \ \ \,
\label{fdefs}
\end{eqnarray}
In addition, the usual continuity equation  is still preserved:
\begin{eqnarray}
 \dot{\rho}_{MG} +3H(\rho_{MG}+p_{MG})=0.
\end{eqnarray}

Variation of the action with respect to the scalar field $\psi$ provides
its evolution equation:
\begin{eqnarray}
\label{psievol}
&&\ddot{\psi}+3H\dot{\psi}+\frac{d W}{d \psi}
+\frac{d V}{d\psi}
\left\{\left(\frac{a_{ref}}{a}-1\right)
[f_3(a)+f_1(a)]+3\dot{b}
f_2(a)\right\}  =0 .\ \ \ \ \ \
\end{eqnarray}
Additionally, variation  with respect to
$b$ provides the constraint equation
\begin{eqnarray}
\label{constraint}
 V(\psi)Hf_1(a)+\dot{V}(\psi)f_2(a) =0 ,
\end{eqnarray}
which, as it was shown in  \cite{Saridakis:2012jy}, using  (\ref{fdefs}) in
general
leads to
\begin{eqnarray}
\label{constraint2}
&&V(\psi(\tau)) = V_0\,e^{-\int \frac{f_1}{af_2}\ud a}
 =
\frac{V_0}{(a-a_{ref})[
\alpha_4a_{ref}^2-(3\alpha_3+2\alpha_4)aa_{ref}+(3+3\alpha_3+\alpha_4)a^2 ]
},\ \ \ \ \ \ \ \
\end{eqnarray}
This relation restricts radically the coupling-potential $V(\psi)$
\footnote{Note that in standard massive gravity this equation imposes
strong constraints on $b(\tau)$
 \cite{D'Amico:2011jj,Gumrukcuoglu:2011ew,Langlois:2012hk}, but in the
present construction the extra freedom brings these constraints on
$V(\psi)$, leaving $b(\tau)$ free  \cite{Saridakis:2012jy}.} and
can be very
helpful since it offers the behavior of $V(\psi(\tau))$ without the need
to find explicitly the solution $\psi(\tau)$ [which obviously is
consistent with this $V(\psi(\tau))$].  Lastly, the equations close by
considering the matter evolution
equation
$\dot{\rho}_m +3H(\rho_m+p_m)=0$.

Finally, we mention that from the above expressions we observe that
$a_{ref}$ plays the role of a reference scale factor that can be arbitrary.
One could still worry about the fact that $V(\psi)$ in general becomes
negative for $a<a_{ref}$, which can be pushed to very small  but still
non-zero values, however the bouncing behavior of the present work offers a
solution to this problem, since setting the bounce scale factor to be
larger than  $a_{ref}$   ensures that the graviton mass square will be
always positive. For simplicity throughout this  work we set
$a_{ref}=1$.

\section{Bouncing and Cyclic solutions}
\label{bouncingsol}

The above cosmological scenario proves to exhibit very interesting
behavior. In particular, the form of the coupling-potential
(\ref{constraint2}) implies that in an expanding universe $V(\psi(\tau))$
always goes to zero at late times, and thus the scenario at hand always
gives
the standard quintessence scenario  \cite{Saridakis:2012jy}. The only case
that this
will not happen is if the scalar field dynamics is so effective that it
will change the universe evolution from expansion to contraction. In this
case $V(\psi(\tau))$ will start increasing and at some point it can
trigger a nonsingular bounce. A successive sequence of bounces and
turnarounds offers the cyclic cosmology. In this section we analytically
explore these possibilities.

Whether a universe is expanding or contracting depends on the positivity of
the Hubble parameter, that is, in the contracting phase $H$ is negative
while in the expanding one it is positive. By making use of the continuity
equations it follows that at the bounce and turnaround points $H=0$,
however at and around the bounce we have $\dot H> 0$, while at and around
the turnaround we have $\dot H < 0$. If a bounce and a turnaround can be
realized in a given cosmological scenario, then the aforementioned general
requirements for $H$ must be fulfilled.

Interestingly enough, observing the form of the two Friedmann equations
(\ref{Fr1}), (\ref{Fr2}) along with
(\ref{rhomg}), (\ref{pmg}), (\ref{constraint2}), we can see that the above
requirements can be easily fulfilled in the present scenario, for suitable
potentials $W(\psi)$ and/or suitable forms
for $b(\tau)$. In particular, the coupling-potential $V(\psi)$ can trigger
the bounce and the turnaround, and actually it is this term that can cause
the necessary null energy condition violation, as it was mentioned in
 \cite{Saridakis:2012jy}, where this term could lead the effective dark
energy caused
by the gravitational modification to exhibit phantom behavior.

In summary, cosmology governed by extended massive gravity has a large
freedom to fulfill
the above bounce and cyclicity requirements. In particular, for a given
$b(\tau)$ we can adjust the potential $W(\psi)$ in order to obtain a
desired scale-factor evolution, or alternatively for a given $W(\psi)$ we
can adjust $b(\tau)$. In the following subsections we examine these two
cases separately.

\subsection{Known $b(\tau)$}
\label{caseb}

Suppose we determine $b(\tau)$ at will. Let us first start from the
desired result, that is we impose a known form of the scale factor
$a(\tau)$ possessing a bouncing or oscillatory behavior. In this case
both $H(\tau)$ and $\dot{H}(\tau)$ are straightforwardly known.
Therefore, we can use the Friedmann equations (\ref{Fr1}), (\ref{Fr2})
together with (\ref{rhomg}),(\ref{pmg}), in order to extract the
relations for $\psi(\tau)$ (through $\dot{\psi}(\tau)$) and $W(\tau)$,
obtaining
 \begin{eqnarray}
 \label{psidot}
 &&\psi(\tau) = \pm \int^\tau d\tau' \Big\{-2M_P^2\dot
H(\tau')- \rho_m(a(\tau'))-p_m(a(\tau'))
 \nonumber\\
&& \ \ \ \ \ \ \ \ \ \ \ \ \ \ \ \ \ \ \ \ \  \ \ \ \ \ \ \ \
\ \ \ \ \ \ \ \
\ \ \ \ \ +V(\psi(\tau'))\left[\dot{b}(\tau')-\frac{1}{a(\tau')}
\right]f_1(a(\tau')) \Big\}^{\frac{1}{2}}  ~, \\
 \label{Wt}
 &&W(\tau) =
M_P^2\left[3H^2(\tau)+\dot{H}(\tau)\right]+\frac{p_m(a(\tau))}{2}-\frac{
\rho_m(a(\tau))}{2} \nonumber\\
&& \ \ \ \ \ \ \ \ \ \ \ \ \ \ \ \ \ \ \ \ \  \ \ \ \ \ \ \ \
\ \ \ \ 
\ \ \ \ \, -V(\psi(\tau))\left\{
f_4(a(\tau))+\frac{f_1(a(\tau))}{2}\left[\dot{b}
(\tau)+\frac{1}{a(\tau)}\right] \right\}
~,
\end{eqnarray}
where $\rho_m(a(\tau))$ and $p(a(\tau))$ are known as long as the matter
equation-of-state parameter $w_m\equiv p_m/\rho_m$ is determined.
Then, eliminating time between these two expressions we can extract the
explicit form of the potential $W(\psi)$, which is the one that generates
the desired $a(\tau)$-form. Finally, we mention that the basic requirement
for
the above procedure to be valid and a bounce or cyclic behavior to be
possible, is that the $a(\tau)$ and $b(\tau)$ ansatzes  and the parameter
choices must lead to a positive $\dot{\psi}^2(\tau)$.

We now proceed to a specific, simple, but quite general example. Firstly,
we consider $b(\tau)$ to have the simple form
\begin{equation} \label{btau}
b(\tau)=\tau,
\end{equation}
and thus $\dot{b}=1$ in the above expressions (in general we could add a
coefficient $b_0$ in the above ansatz, but for simplicity we set it equal
to 1). Now, in order to get an
idea of what kinds of potentials $W(\psi)$ could lead to cyclic behavior,
we perform a test-procedure starting from the desired result, that is we
assume a cyclic universe with an oscillatory scale factor of
the form
\begin{equation} \label{atosc}
a(\tau)=A\sin(\omega \tau)+a_c,
\end{equation}
where the
non-zero constant $a_c$ is inserted in order to avoid the
singularity. Thus, $\tau$
varies between $-\infty$ and $+\infty$, with $\tau=0$ being a
specific moment without any particular physical meaning. Finally,
note that the bounce occurs at $a_{B}=a_c-A$. From (\ref{atosc})
we straightforwardly   find $H(\tau)$ and $\dot{H}(\tau)$,
and thus substitution into (\ref{constraint2}) and then to
(\ref{psidot}) and (\ref{Wt}) gives the corresponding
expressions for $\psi(\tau)$ and $W(\tau)$.

\begin{figure}[ht]
\begin{center}
\mbox{\epsfig{figure=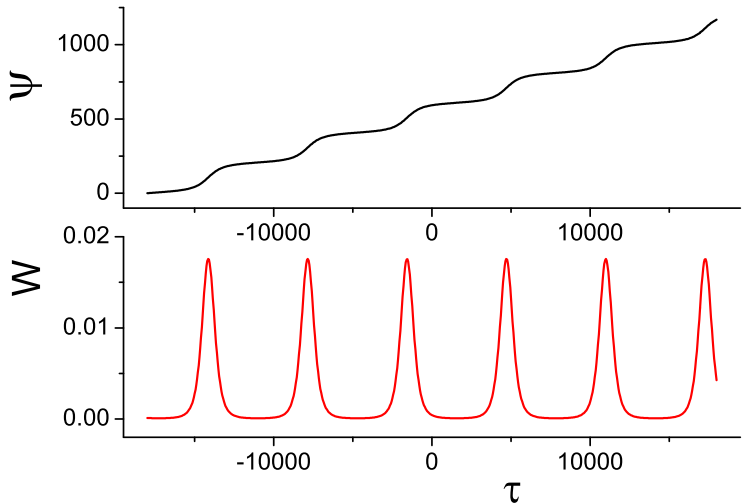,width=11.cm,angle=0}} \caption{{\it The
evolution of $\psi(\tau)$ and $W(\tau)$ as functions of time in the case
where $b(\tau)=\tau$ and $a(\tau)=A\sin(\omega \tau)+a_c$, with $A=5$,
$\omega=0.001$, and $a_c=7$.  In numerical
elaboration we use $\alpha_3=3$, $\alpha_4=1$, $V_0=0.2$, $\rho_{mB}=0.01$,
and $M_P=1$. All dimensional parameters are normalized in unit of $M_P$. 
}}
\label{Wpsitau1}
\end{center}
\end{figure}

\begin{figure}[ht]
\begin{center}
\mbox{\epsfig{figure=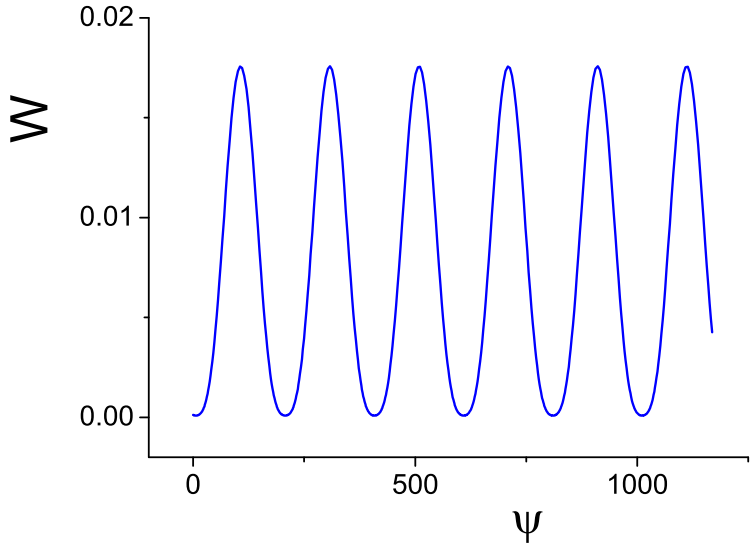,width=11.cm,angle=0}} \caption{{\it The
reconstructed potential $W(\psi)$ in the case
where $b(\tau)=\tau$ and $a(\tau)=A\sin(\omega \tau)+a_c$, with $A=5$,
$\omega=0.001$, and $a_c=7$.  In numerical
elaboration we use $\alpha_3=3$, $\alpha_4=1$, $V_0=0.2$, $\rho_{mB}=0.01$,
and $M_P=1$. All dimensional parameters are normalized in unit of $M_P$. 
}}
\label{Wpsi}
\end{center}
\end{figure}

In order to provide a more transparent picture of the obtained
cosmological behavior, in Fig.~\ref{Wpsitau1} we present the evolution
of
$\psi(\tau)$ and $W(\tau)$ for the scale factor (\ref{atosc}) with $A=5$,
$\omega=0.001$ and $a_c=7$,
assuming also
dust matter  ($w_m=0$), that is $p_m=0$ and
$\rho_m(a)=\rho_{mB}a_{B}^3/a^3$, with $\rho_{mB}$ the energy density at
the
bounce.
Thus, eliminating time between $\psi(\tau)$ and $W(\tau)$ allows
us to re-construct the corresponding expression for $W(\psi)$, shown in
Fig.~\ref{Wpsi}.
From these figures we observe that an oscillating and
non-singular scale factor can be generated by an oscillatory
scalar potential $W(\psi)$. This $W(\psi)$-form was expected, since a
non-oscillatory $W(\psi)$ would be difficult to
produce an infinitely oscillating scale factor. Finally, we mention that
although we have presented a specific example, we can
straightforwardly repeat the described procedure imposing an
arbitrary oscillating form for the scale factor.

The aforementioned bottom-to-top approach was enlightening about
the form of the scalar potential that leads to a cyclic
cosmological behavior. Therefore, one can perform the above
procedure restoring cause and effect, that is starting from a specific
oscillatory $W(\psi)$ and resulting to an oscillatory $a(\tau)$. In
particular, (\ref{psidot}) is written in a compact form as
$\dot{\psi}^2(\tau)=P_1(a,\dot{a},\ddot{a})$ and similarly
(\ref{Wt}) as $ W(\tau)=P_2(a,\dot{a},\ddot{a})$. Thus, we can
invert  the known form of $W(\psi)\equiv W(\psi(\tau))$ acquiring
$\psi(\tau)=W^{\{-1\}}\left(P_2(a,\dot{a},\ddot{a})\right)$,   and then
$\dot{\psi}^2(\tau)=\left\{\frac{d}{d\tau}\left[W^{\{-1\}}\left(P_2(a,\dot{
a}, \ddot{a})\right)\right]\right\}^2$. Therefore, the scale factor arises
as a solution of the differential equation
\begin{equation} \label{Diffeq}
P_1(a,\dot{a},\ddot{a})=\left\{\frac{d}{d\tau}\left[W^{\{-1\}}\left(P_2(a,
\dot {a},\ddot{a})\right)\right]\right\}^2.
\end{equation}

As a specific example we consider the simple form
\begin{equation} \label{Wtspec}
W(\psi)=W_0\sin(\omega_W\, \psi)+W_c.
\end{equation}
In this case
$\psi=\frac{1}{\omega_W}\sin^{-1}\left[\frac{W(\psi(\tau))-W_c}{W_0}
\right]$,
where $W(\psi(\tau))\equiv W(\tau)=P_2(a,\dot{a},\ddot{a})$  with
$P_2(a,\dot{a},\ddot{a})$ the right hand side of expression
(\ref{Wt}). Therefore, the differentiation leads to:
\begin{equation}
\label{psidotadta}
\dot{\psi}(\tau)=\frac{1}{W_0\,\omega_W}\frac{1}{\sqrt{1-\left[\frac{P_2(a,
\dot{a},\ddot{a})-W_c}{W_0}\right]^2}}
\,\frac{d}{d\tau}\left[P_2(a,\dot{a},\ddot{a})\right]
\end{equation}
and thus we obtain
\begin{equation} \label{Diffeq1}
P_1(a,\dot{a},\ddot{a})=\left\{\frac{1}{W_0\,\omega_W}\frac{1}{\sqrt{
1-\left[\frac{P_2(a,\dot{a},\ddot{a})-W_c}{W_0}\right]^2}}
\,\frac{d}{d\tau}\left[P_2(a,\dot{a},\ddot{a})\right]\right\}^2,
\end{equation}
where as we have mentioned $P_1(a,\dot{a},\ddot{a})$ is the
expression inside the curly brackets in
(\ref{psidot}).

\begin{figure}[ht]
\begin{center}
\mbox{\epsfig{figure=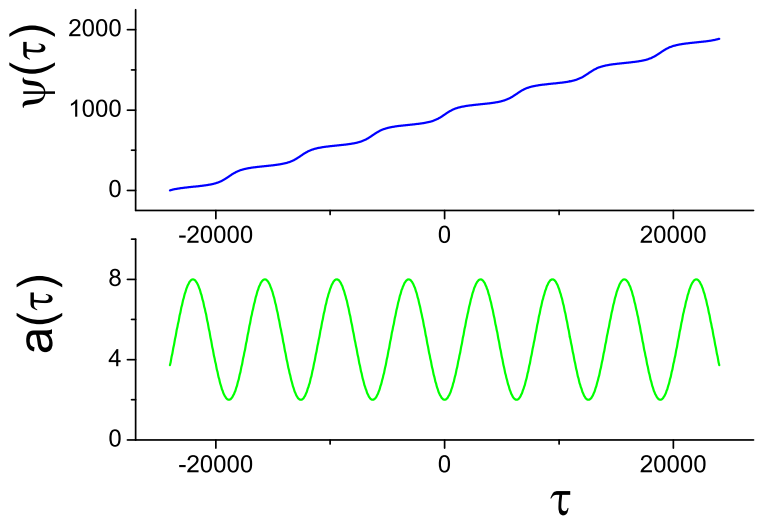,width=11.cm,angle=0}} \caption{{\it The
evolution of the scalar field $\psi(\tau)$ (upper graph) and of the scale
factor (lower graph), for $b(\tau)=\tau$ and for the scalar-field potential
$W(\psi)=W_0\sin(\omega_W\, \psi)+W_c$ with $W_0=0.01$, $\omega_W=0.025$
and
$W_c=0.01$. In numerical elaboration we use $\alpha_3=3$, $\alpha_4=1$,
$V_0=0.2$, $\rho_{mB}=0.01$, and $M_P=1$.
All dimensional parameters are normalized in unit of $M_P$. }}
\label{at1}
\end{center}
\end{figure}

Differential equation (\ref{Diffeq1}) cannot be solved
analytically, but it can be easily elaborated numerically, and moreover
inserting its solution for $a(\tau)$ into (\ref{psidotadta}) we obtain
$\psi(\tau)$ too. In Fig.~\ref{at1} we depict the corresponding solutions
for $a(\tau)$ and   $\psi(\tau)$, for the ansatz (\ref{Wtspec}) with
$W_0=0.01$, $\omega_W=0.025$ and $W_c=0.01$,  for $\alpha_3=3$,
$\alpha_4=1$,
$V_0=0.2$,
$\rho_{mB}=0.01$, $M_P=1$,  in $M_P^2$-units (the
potential parameters have been chosen in order to acquire a cyclic
universe with $a_{B}\approx2$ at the bounce, similarly to the previous
example).
In conclusion, we verified that an oscillatory scalar-field potential
can lead to a cyclic universe.

We close this subsection by mentioning that here $b(\tau)$ was considered
by hand and then we determined the potential $W(\psi)$
 in order to obtain cyclic behavior. The only requirement for the above
analysis to be valid is to obtain a non-negative $\dot{\psi}^2$ in
(\ref{psidot}). If this is not possible we deduce that the corresponding
parameter-choice  and/or
the imposed $b(\tau)$-ansatz  cannot lead to a cyclic solution. However,
note that
there are $b(\tau)$ ansatzes that cannot lead
to cyclicity independently of the parameter choice (for
$\alpha_3$,$\alpha_4$,$V_0$,$\rho_{mB}$,$w_m$). In particular, using the
forms $b(\tau)=a(\tau)$ and
$b(\tau)=\ln a(\tau)$ (which in  \cite{Saridakis:2012jy} it was shown to
lead to
phantom behavior) we could not find any parameter choice that could lead
to non-negative  $\dot{\psi}^2$ and therefore to a cyclic behavior. Thus,
cyclicity cannot be obtained for every $b(\tau)$-form.

\subsection{Known $W(\psi)$}
\label{caseW}

Suppose we determine $W(\psi)$ at will, and similarly to the previous
subsection as a first test-procedure  we impose a
known scale factor $a(\tau)$ possessing  oscillatory
behavior, that is both $H(\tau)$ and $\dot{H}(\tau)$ are
known. In this case eliminating $\dot{\psi}$ between the two Friedmann
equations
(\ref{Fr1}),(\ref{Fr2}) (with $a_{ref}=1$) straightforwardly gives
\begin{eqnarray}
\label{bdotsol}
&& \dot{b}(\tau)=\frac{1}{a(\tau)}+\left[
V(a(\tau))f_1(\tau)\right]^{-1}\left\{2M_P^2\left[\dot{H}
(\tau)+3H^2(\tau)\right]
-\rho_m(a(\tau))+p_m(a(\tau))\right.\nonumber\\
&&\left. \ \ \ \ \ \ \ \ \ \ \ \ \ \ \ \
-2W(\psi(\tau))-
2V(a(\tau))\left[\frac{1}{a(\tau)}-1\right]\left[
f_3(a(\tau))+f_1(a(\tau))\right ]\right \}
  \equiv
P_3(\psi,\tau),\ \ \ \ \
\end{eqnarray}
where $V(a(\tau))$ is given by (\ref{constraint2}),
and thus substitution to the scalar-field evolution equation
(\ref{psievol}), along with the chain rule
$\frac{dV}{d\psi}=\frac{dV}{da}\frac{\dot{a}}{\dot{\psi}}$, leads to a
differential equation for $\psi$  of the form
\begin{equation}
\label{psievol2}
 \ddot{\psi}+3H\dot{\psi}+\frac{d W}{d \psi}
+\frac{dV}{da}\frac{\dot{a}}{\dot{\psi}}
\left\{\left(\frac{1}{a}-1\right)
[f_3(a)+f_1(a)]
+3P_3(\psi,\tau)
f_2(a)\right\} \equiv P_4(\psi,\dot{\psi},\ddot{\psi},t) =0,
\
\end{equation}
that can be numerically solved,
leading to $\psi(\tau)$. Thus, inserting
$\psi(\tau)$ back to (\ref{bdotsol}) gives a simple differential equation
of the form $\dot{b}=P_5(\tau)$ that can be easily solved as
\begin{eqnarray}
\label{bsolf}
b(\tau)=\int^\tau P_5(\tau')d\tau'.
\end{eqnarray}

\begin{figure}[ht]
\begin{center}
\mbox{\epsfig{figure=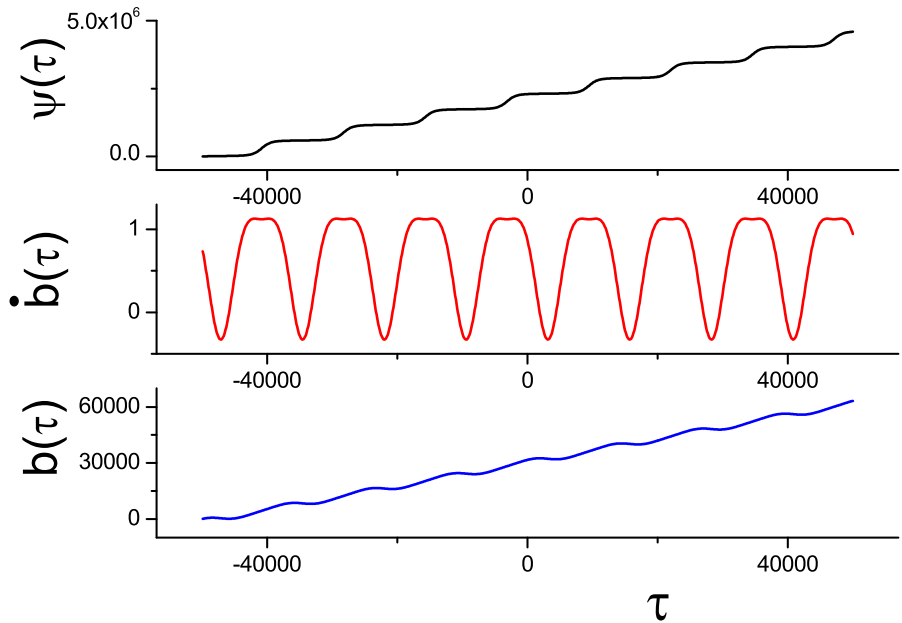 ,width=13.cm,angle=0}} \caption{{\it
The evolution of the scalar field $\psi(\tau)$ (upper graph), of the
time-derivative of the St\"{u}ckelberg-scalar function $\dot{b}(\tau)$
(middle graph) and of the St\"{u}ckelberg-scalar function ${b}(\tau)$
(lower graph), in the case where $W(\psi)=W_0$ with $W_0=0.001$ and
$a(\tau)=A\sin(\omega \tau)+a_c$, with  $A=3$,
$\omega=0.0005$, and $a_c=5$.
In numerical elaboration we use $\alpha_3=10$, $\alpha_4=10$, $V_0=0.6$,
$\rho_{mB}=0.1$, and $M_P=10$.
All dimensional parameters are normalized in unit of $M_P$. }}
\label{btaupsitau1}
\end{center}
\end{figure}

As a specific example we consider the simplest case in which $W(\psi)$ is
constant, namely
\begin{equation} \label{Wpsi1}
W(\psi)=W_0.
\end{equation}
 Similarly to the previous subsection we assume a cyclic
universe with the oscillatory scale factor  (\ref{atosc}), namely
$a(\tau)=A\sin(\omega \tau)+a_c$. In this case from (\ref{bdotsol}) we can
see that $\dot{b}(\tau)$ is analytically known, which simplifies the
procedure. Assuming also
dust matter  ($w_m=0$) and
$\rho_m(a)=\rho_{mB}a_{B}^3/a^3$, with $\rho_{mB}$ the energy density at
the
bounce, we numerically solve the differential equation
(\ref{psievol2}), in order to extract
$\psi(\tau)$. In Fig.~\ref{btaupsitau1} we depict the corresponding
evolution of $\psi(\tau)$, $\dot{b}(\tau)$ and ${b}(\tau)$, and thus we
deduce that this $b(\tau)$ form generates the oscillatory universe.
 As we observe, $\dot{b}(\tau)$ has an oscillatory profile, although not
of a simple form. The fact that $\dot{b}(\tau)$ and
not $b(\tau)$ should be oscillatory in order to produce an oscillatory
scale factor was expected, since in the cosmological equations it is
$\dot{b}$ and not $b$ that appears.

We stress here that the differential equation (\ref{psievol2}) in the
above procedure cannot be solved for every $W(\psi)$, and additionally one
has to suitably choose the model parameters. Thus, although in the previous
subsection one had full control on which cases lead to oscillatory
solutions (namely those that give a positive $\dot{\psi}^2(\tau)$ in
(\ref{psidot})), in the present subsection one cannot easily see which
$W(\psi)$ ansatzes and/or which parameter choices can lead to cyclic
solutions.

Finally, since from the above bottom-to-top approach we deduced that in
order to get an oscillatory universe we must use an oscillatory
$\dot{b}(\tau)$, in the following we restore cause and effect and we
impose such a $\dot{b}(\tau)$ as an input, extracting the resulting
$\psi(\tau)$ and $a(\tau)$ from the solutions of the cosmological
equations.
Although this procedure is possible in general, in the following we apply
it in the simple example of a constant $W(\psi)=W_0$ of (\ref{Wpsi1}). In
this case, and with a given $\dot{b}(\tau)$, equation (\ref{bdotsol}) is
a simple differential equation for $a(\tau)$, which can be numerically
solved. Then, inserting this $a(\tau)$ back in (\ref{psidot}), we find
also the solution for $\psi(\tau)$. Similarly to the previous subsection,
the only requirement for this procedure to work is to obtain a positive
$\dot{\psi}^2(\tau)$ .

\begin{figure}[ht]
\begin{center}
\mbox{\epsfig{figure=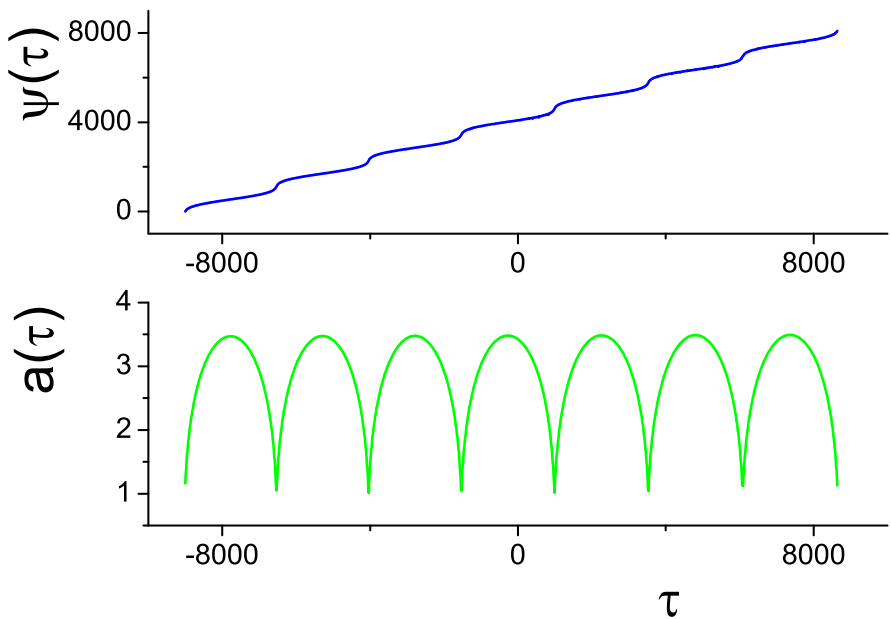 ,width=13.cm,angle=0}} \caption{{\it
The evolution of the scalar field $\psi(\tau)$ (upper graph) and of the scale
factor (lower graph), in the case where $W(\psi)=W_0$ and
$\dot{b}(\tau)=B\sin(\omega_B \tau)+b_c$,
with $W_0=0.001$, $B=1$, $\omega_b=10^{-6}$ and
$b_c=2$. In numerical elaboration we use $\alpha_3=4$, $\alpha_4=4$,
$V_0=1$, $\rho_{mB}=0.0005$, and $M_P=100$.
All dimensional parameters are normalized in unit of $M_P$. }}
\label{ataupsitau1}
\end{center}
\end{figure}

As a specific example for $\dot{b}(\tau)$ we impose the simple ansatz
\begin{equation} \label{bdansatz}
\dot{b}(\tau)=B\sin(\omega_B \tau)+b_c,
\end{equation}
that is we impose a St\"{u}ckelberg-scalar function of the form
\begin{equation} \label{bansatz}
b(\tau)=-\frac{B}{\omega_B}\cos(\omega_B \tau)+b_c \tau+c,
\end{equation}
with $c$ an integration constant. We then solve numerically
(\ref{bdotsol}) for $a(\tau)$ and (\ref{psidot}) for $\psi(\tau)$ and in
  Fig.~\ref{ataupsitau1} we depict the corresponding
results.
Indeed, the St\"{u}ckelberg-scalar function (\ref{bansatz})
generates an oscillatory universe, although not of a simple form since the
turnarounds are softer than the bounces. However, we mention that one
needs to carefully select the model parameters in order to   obtain an
oscillatory scale factor under the $b(\tau)$-ansatz (\ref{bansatz}),
however he has control of the procedure, since he needs to fulfill the
requirement of $\dot{\psi}^2(\tau)$ to be positive.

\section{Discussion}
\label{Discussion}

In the present work we investigated bouncing and cyclic cosmological
behaviors in a universe governed by extended massive gravity, in which the
graviton mass has been promoted to a function of an extra scalar field.
This model has additional freedom comparing to usual massive gravity, and
thus it leads to significantly different and richer behavior
\cite{Huang:2012pe,Saridakis:2012jy}. In particular, although the scenario
at late times tends to coincide with standard quintessence, at early and
intermediate times the effective graviton mass can be large and thus play
a crucial role in the universe evolution. Amongst others, the capability of
the scenario to lead to violation of the Null Energy Condition can lead to
a bounce or a turnaround, the successive sequence of which can naturally
give rise to cyclic cosmology.

Extended nonlinear massive gravity has an enhanced freedom (apart from the
extra
scalar field one can see that the St\"{u}ckelberg-scalar function $b(\tau)$
is not constrained as in usual massive gravity and it can be free),
therefore it can drive cyclicity with two different mechanisms. The first
is to impose an
arbitrary St\"{u}ckelberg-scalar function $b(\tau)$ and suitably choose the
usual scalar field potential $W(\psi)$ in order to obtain a cyclic
behavior. It proves that one should use an oscillating  $W(\psi)$ as
expected, and the only requirement for this procedure to hold is to obtain
a positive $\dot{\psi}^2(\tau)$. Thus, this is not possible for every
$b(\tau)$ form, that is not every  $b(\tau)$ can be
consistent with cyclicity.

The second mechanism to drive cyclic behavior is exactly the
St\"{u}ckelberg-scalar function $b(\tau)$. In particular, imposing an
arbitrary scalar-field potential $W(\psi)$ we suitably choose $b(\tau)$ in
order to obtain a cyclic behavior. It proves that one should use an
oscillating  $\dot{b}(\tau)$ as expected (since $\dot{b}(\tau)$ appears in
the equations and not $b(\tau)$). Similarly to the first mechanism above,
not all scalar potentials are consistent with cyclicity.

A crucial issue in all bouncing and cyclic scenarios is the processing of
perturbations through the bounces. A simple check on the stability of this
type of models is to look at the generalized Higuchi bound derived in
\cite{Fasiello:2012rw} (see also \cite{Grisa:2009yy} for a different
expression in an earlier model). In particular, as an effective theory,
nonlinear massive gravity   is  reliable only when the scale is well
beneath the cut off $\Lambda_3 = (M_P m_g^2)^{1/3}$, where $m_g^2$ is the
 graviton mass square, since upon $\Lambda_3$  helicity 0 mode 
couples strongly to helicity 1 and helicity 2 modes, and thus effective
field
theory breaks down. Therefore, in the present extended scenario one should
compare\footnote{We wish to thank the referee for mentioning this point.}
the various appeared scales, such are $H$,$\dot{H}$,$\ddot{H}$ etc, with
$\Lambda_3 = [M_P V(\psi)]^{1/3}$. Indeed, in all the above examples
$H/\Lambda_3\lesssim10^{-3}$, while $\dot{H}/\Lambda_3\lesssim10^{-5}$, 
 $\ddot{H}/\Lambda_3\lesssim10^{-7}$ etc, and thus the scenario is
reliable. On the other hand note that $V(\psi)$ is always much smaller
than $M_P^2$ ($V(\psi)/M_P^2\lesssim10^{-3}$ and
$V_0/M_P^2\lesssim10^{-1}$),
which is an additional requirement for the robustness of the scenario. 
Therefore, at this level, we can conclude that our model is well behaved
when the perturbations pass through the bouncing points. However, we should
notice that since a cosmic scalar is introduced to drive the graviton mass
varying along background evolution, the stability issue arisen from this
scalar field ought to be taken into account in a global analysis. Such a
complete perturbation analysis of the extended nonlinear massive gravity
lies beyond the scope of the present work and it is left for future
investigation.

\vskip .2in \noindent {\large{{\bf {Acknowledgments}}}}

We wish to thank S. Dutta, Q-G. Huang, Y.-S, Piao, S.-Y. Zhou and an
anonymous referee, for
useful comments.
The work of CYF is supported in part by the Cosmology Initiative in Arizona
State University. The work of GCX is supported in part by the Summer Research
Assistantship from the Graduate School of the University of Mississippi and
in part by the Cosmology Initiative in Arizona State University.
The research of ENS is implemented within the framework of the Action
Supporting Postdoctoral Researchers of the Operational Program
``Education and Lifelong Learning'' (Actions Beneficiary: General
Secretariat for Research and Technology), and is co-financed by
the European Social Fund (ESF) and the Greek State.\\

\providecommand{\href}[2]{#2}

\begingroup

\raggedright

\endgroup

\end{document}